\documentclass[12pt]{article}
\usepackage{graphicx}
\usepackage{amsmath,amssymb}

\begin{document}
\begin{center}
\large
{\large \bf  Boltzmann entropy and chaos in a large assembly}\\
{\bf of weakly interacting systems }\\
\normalsize
 \vspace{0.3in}
{Massimo Falcioni$^1$, Luigi Palatella$^1$, Simone Pigolotti$^1$, \\
Lamberto Rondoni$^2$ and Angelo Vulpiani$^{1,3}$}\\
 \vspace{0.1in}
{$^1$ Dipartimento di Fisica and Center for Statistical 
Mechanics and Complexity - INFM,  Universit\`a di Roma "La Sapienza", 
P.le A.Moro 2, Rome 00185, Italy\\}
\vspace{0.1in}
{$^2$ Dipartimento di Matematica and INFM, Politecnico di Torino
Corso Duca degli Abruzzi 24, 10129 Torino - Italy \\}
\vspace{0.1in}
{$^3$ INFN, Sezione di Roma ``La Sapienza''}

\vspace{0.2in}
\end{center}
\date{\today}

\noindent
{\it Corresponding author:} Lamberto Rondoni\\
\makebox[23mm]{}tel:  +39 + 011 + 5647533\\
\makebox[23mm]{}fax: +39 + 011 + 5647599  \\
\makebox[23mm]{}e-mail: lamberto.rondoni@polito.it

\newpage

\begin{abstract}
We introduce a high dimensional symplectic map, modeling a large
system consisting of weakly interacting chaotic subsystems, as a toy
model to analyze the interplay between single-particle chaotic dynamics
and particles interactions in thermodynamic systems. We study the
growth with time of the Boltzmann entropy, $S_B$, in this system as a
function of the coarse graining resolution. We show that a
characteristic scale emerges, and that the behavior of $S_B$ vs $t$, at
variance with the Gibbs entropy, does not depend on the coarse
graining resolution, as far as it is finer than this scale. The
interaction among particles is crucial to achieve this result, while  
the rate of entropy growth depends essentially on the single-particle
chaotic dynamics (for $t$ not too small). It is possible to interpret
the basic features of the dynamics in terms of a suitable Markov
approximation.
\end{abstract}

\vspace{0.1in}

\section{Introduction}

Statistical mechanics was founded by Maxwell, Boltzmann and Gibbs for
systems with a very large number of particles. Given their different
approaches, Boltzmann and Gibbs are often considered as the champions
of two different points of view about statistical mechanics. This
common {\em vulgata} takes Gibbs as the founder of the ensemble
approach, and Boltzmann as the promoter of a dynamical theory based on
the ergodic hypothesis; according to this tradition, modern textbooks
\cite{huang} use Gibbs's terminology for the ensembles
(i.e. microcanonical, canonical and grandcanonical).  Actually, both
ergodicity and ensembles are Boltzmann's inventions; more detailed
discussions about Boltzmann, Gibbs, and the origin of the statistical
mechanics are contained in the absolutely recommendable book by
Cercignani \cite{cercignani}, and the papers by Klein \cite{klein},
Lebowitz \cite{lebowitz_PT}, and Gallavotti \cite{gallavotti}.

The Gibbs' approach to statistical mechanics uses the general concept
of ensembles without any reference to the precise origin of such an
idea.  The relevant literature is huge \cite{jaynes,klein} : sometimes
the ensembles are derived from philosophical considerations, other
times from information-theoretic, and even from anthropocentric
considerations. In Van Kampen's words, sometimes this produces {\it a
slightly mystic aura}, which diverts the attention away from the
physical meaning of the ensemble theory \cite{vankampen}.

Ergodic theory begins with Boltzmann's attempt to justify the
determination of average values in kinetic theory
\cite{ehrenfest,gallavotti}.  Macroscopic systems contain a very large
number (of the order of the Avogadro's number) of particles; this
implies the practical necessity of a statistical description.  Since
in any macroscopic observation the time scale is much larger than the
microscopic time scale, over which the molecular changes take place,
an experimental measurement is actually the result of an observation
on a single system, during which it goes through a very large number
of microscopic states. Thus, one may assume that the outcome of the
measurement of an observable $A({\bf X})$ is an average performed over
all these states:
\begin{equation}
\overline{A}^T({\bf X}_0) = {1 \over T} \int_{t_0}^{t_0+T} 
A({\bf X}(t;{\bf X}_0)) {\rm d} t \, ,
\label{Amedio}
\end{equation}
where the vector
\begin{equation}
{\bf X}(t;{\bf X}_0) \equiv ({\bf q}_1 (t;{\bf X}_0) , \dots , {\bf
 q}_N (t;{\bf X}_0), {\bf p}_1 (t;{\bf X}_0),\dots , {\bf p}_N (t;{\bf
 X}_0) )
\end{equation}
lies in the $6\, N$ - dimensional phase space $\Gamma$, gives the
state of the system at time $t$ and, in general, depends on the
initial state ${\bf X}_0$. According to the usual notation, the
components of ${\bf X}$, $ {\bf q}_i $ and $ {\bf p}_i $, indicate the
position and momentum vectors of the $i$-th particle.

Obviously, it is impossible to obtain the complete description ${\bf
X}_0$ of the microscopic state of the system at the initial time, for
such high-dimensional systems as those relevant in statistical
mechanics.  This means that, if $\overline{A}^T$ depends too strongly
on the initial condition, no statistical predictions can be made,
independently of how difficult it may be to integrate the equations of
motion.  The ergodic hypothesis allows us to overcome this obstacle,
stating that for sufficiently large $T$, the average $\overline{A}^T$
depends only on the system energy, in the sense that it takes the same
values for {\em almost all}\footnote{Here, {\em almost all} has a
technical meaning, i.e.\ it
refers to all trajectories, except those which originate in a set of
vanishing phase space volume.}
the trajectories on the given
constant energy surface. This allows us to identify the time average
with a phase space average:
\begin{equation}
\label{ergo1}
\overline{A} \equiv \lim_{T \to \infty}  {1 \over T} \int_{t_0}^{t_0+T} 
A({\bf X}(t;{\bf X}_0)) {\rm d} t = \int A({\bf X}) P_{mc}({\bf X} ) 
{\rm d} {\bf X} \equiv \langle A \rangle \, 
\end{equation}
where $P_{mc}$ is the microcanonical probability density, except for a
negligible set of initial conditions.  The validity of such an
equality eliminates the necessity of determining the initial state of
the system, and of solving the Hamilton's equations. Whether
(\ref{ergo1}) is valid or not, i.e.\ whether it is possible to
substitute the temporal average with an average in phase space,
constitutes the essence of the ergodic problem in physics.  The
relevance of this issue is evident: let the statistical properties of
a large isolated system be properly described by the microcanonical
ensemble, then it is not difficult to show that a small part of the
given system, which is still large at the microscopic scale, is
described by the canonical ensemble.

We stress that, in this context, the ensemble is just a useful
mathematical tool, and one should not forget that thermodynamics, as a
physical theory, is developed to describe the properties of {\em
single} systems, made of many microscopic, interacting parts.
Although some may find hard to distinguish between ``thermodynamic
systems'' and other systems of interest in statistical physics, in
this paper, when referring to a thermodynamic system, we have in mind
a system that is characterized by macroscopic properties such as its
temperature, while many other, equally interesting physical systems,
do not even need to have a temperature. The distinction makes sense
because different laws describe the behaviour of the different
systems.

To justify eq.\ (\ref{ergo1}) in physical systems, Khinchin's point of
view clearly states that the success of statistical mechanics
techniques is mainly due to the many degrees of freedom, while the
underlying dynamics play only a minor role. In his celebrated book
{\it Mathematical Foundations of Statistical Mechanics}
\cite{khinchin}, he presents some important results on the ergodic
problem which do not rest on the metrical transitivity of the
dynamics, which (\ref{ergo1}) requires.  Khinchin's idea on the
ergodic problem in statistical mechanics is that one should profit
from the fact that macroscopic systems have a large number of degrees
of freedom, and that the physical observables for which (\ref{ergo1})
is required to hold are not generic regular functions, as in the
dynamical systems approach to ergodicity. In other words, taking into
account the large number of degrees of freedom should be enough to
obtain the validity of the equivalence between time averages and
ensemble averages for the small class of relevant observables.
Moreover, it is fair for physical purposes to accept the failure of
the ergodic hypothesis for a set of initial conditions ${\bf X}_0$
which is a negligible fraction of the phase space, or that becomes
negligible when $N \to \infty$. In a nutshell, Khinchin's point of
view is that statistical mechanics works independently of the
mathematical validity of the dynamical systems notion of ergodicity,
because the physically relevant observables are practically constant,
except in an irrelevant region of the constant energy surface. This
makes marginal the role of the dynamics, for equilibrium states, and
the existence of good statistical properties is mainly due to the
large number of degrees of freedom, i.e.\ to $N \gg 1$.

The ergodic approach, whether in its strong or its weak version, is a
natural way to introduce probabilistic concepts in a deterministic
context: in experimental situations one deals with a unique system
with many degrees of freedom, whose microstates explore with given
frequency the different regions of the phase space. Consequently, the
ergodic hypothesis seems to provide the appropriate tools for the
equilibrium statistical mechanics to be founded on physical grounds,
using a frequentistic interpretation of the probability distributions.

The other way of connecting probability and dynamics (which does not
contrast with Boltzmann's point of view) is to take into account the
unavoidable uncertainty on the initial conditions. This approach,
originally due to Maxwell, considers that there are {\it a great many
systems the properties of which are the same, and that each of these
is set in motion with a different set values for the coordinates and
momenta} \cite{maxwell}.

The discovery of deterministic chaos, implying that even deterministic
systems with a few degrees of freedom may present some statistical
features typical of probabilistic evolutions, forced the physicists to
reconsider from a new perspective the foundations of statistical
mechanics \cite{bricmont}.  Moreover one may also find useful to
define, formally, ``thermodynamic'' system any system with ergodic
dynamics such that (\ref{ergo1}) applies.  In this paper we will
consider only systems with many degrees of freedom, such that,
usually, the thermodynamic observables have negligible fluctuations.

The situation is much more intricate for the non equilibrium problems
than for equilibrium ones, and even after many years of debate, there
is no general agreement about the fundamental ingredients needed for
statistical mechanics to hold.

The aim of this paper is to give a contribution for the understanding
of the role of deterministic chaos, and of coarse-graining procedures,
in non equilibrium statistical mechanics. In particular the paper
concerns the problem of the growth of entropy. These issues are the
source of heated debates, on which only partial agreement has been
reached within the statistical mechanics community. Recent examples of
this are given by the debate on irreversibility originated by
Lebowitz's paper \cite{lebowitz_PT}, and the partly contrasting views
on irreversible entropy production reported in
\cite{PG98,TVM01,RC-CR,GDN03}.\footnote{As a matter of fact, the views
expressed in Refs.\cite{TVM01} are not particularly contrasting with
those of Refs.\cite{RC-CR}, as they all require the condition of local
thermodynamic equilibrium, and do the coarse graining in $\mu$-space
rather than in $\Gamma$-space.}

Our results support the view that systems of many noninteracting
chaotic particles may properly describe, for some aspects, highly
rarefied gases, or Knudsen gases in an irregular container, even if a
normal thermodynamic behavior may be absent.\footnote{The lack of
interactions imparts very interesting and useful (cf.\
Ref.\cite{DGM,JR05}) properties to such gases, not their number
density. Therefore, the behaviour of a system of noninteracting
particles, like that of a Knudsen gas, is dominated by the collisions
with the walls of its container, even if its number of particles per
unit volume is high.} This is a very delicate issue: it is known
\cite{grad,lebo-spo} that some form of irreversibility may be found in
systems of non interacting particles. For example, free particles in a
box show homogenization of the spatial coordinates; in irregularly
shaped boxes the distribution of velocity can become isotropic; but,
in the absence of interactions, a relaxation of velocities to a
Maxwellian (that we call, as noted above, a \textit{normal
thermodynamic behaviour}) is impossible.  We find, in fact, that the
$\Gamma$ or $\mu$ space graining required for noninteracting particle
systems to look thermodynamic-like have an equally arbitrary
character, with no physical meaning, and play an essentially identical
role in both spaces.

Differently, we are going to show in this paper that, in the case of
interacting chaotic particles, a characteristic scale naturally
emerges in the $\mu$ space. This may be interpreted as the scale at
which the diffusive, small scale behavior due to the interaction first
smoothes the fragmented structures created by the chaotic dynamics. In
this case, the coarse graining in $\mu$-space shows completely
different features from the coarse graining in $\Gamma$ space: in
particular, the growth of the Boltzmann entropy results independent of
the resolution, as long as the observation scale is finer than the
characteristic scale. We stress that to achieve this result, it is of
fundamental importance that the number of particles must be large,
which is not required for the coarse graining in $\Gamma$-space.

These facts are reflected in the different nonequilibrium behaviours
of the Boltzmann and Gibbs entropies. Indeed, the constancy of the
Gibbs entropy can be consistently accommodated in this framework.  If
one accepts, as we do, that the Gibbs canonical ensemble does yield
the correct equilibrium thermodynamics, then the dynamical invariance
of the Gibbs entropy may provide a simple interpretation of the second
law of thermodynamics \cite{jaynes1}. One considers the ensemble of
points in $\Gamma$ space, that are compatible with the values of the
macroscopic variables $\lbrace A_k \rbrace$, defining the initial
equilibrium thermodynamic state, and that are weighted by the suitable
Gibbs density $\rho (\textbf{X})$. For large systems, the quantity
$H_G(\{\rho\}) = \int \rho(\mathbf{X})\ln \rho(\mathbf{X})\mathrm{d}
\mathbf{X}$ allows to determine a phase space region $R$, whose volume
is $\Delta \Gamma = \exp [- H_G(\{\rho\}]$, containing the
``reasonably probable'' microstates. As a consequence of an adiabatic
change of state, the microstates in $R$ evolve, and after a
sufficiently long time $T$, so that a new equilibrium is attained,
they occupy the new region $R_T$, whose volume, because of the
Liouville equation, is still $\Delta \Gamma$.  However, the quantities
defining the new thermodynamic equilibrium are $\lbrace
\widetilde{A}_k \rbrace$, evolved from $\lbrace A_k \rbrace$.  These
values define the new Gibbs density $\widetilde{\rho} (\textbf{X})$,
with a support essentially contained in a region $\widetilde{R}$ of
volume $\Delta \widetilde{\Gamma} = \exp [-
H_G(\{\widetilde{\rho}\}]$.  Since we accept that, at equilibrium, $S
= - k_{_B} \int \rho(\mathbf{X})\ln \rho(\mathbf{X})\mathrm{d}
\mathbf{X} = - k_{_B} H_G(\{\rho\}) = k_{_B} \ln \Delta \Gamma $, the
second law of thermodynamics requires that $\Delta\widetilde{\Gamma}
\geq \Delta \Gamma$. What is important to note here is that
$\widetilde{\rho} (\textbf{X})$, and therefore
$\Delta\widetilde{\Gamma}$, is not determined by the evolution of the
initial $\rho (\textbf{X})$, it is constructed by the evolved values
of the macrovariables $\lbrace A_k \rbrace$. A demonstration of the
second law in statistical mechanics, in this framework, is equivalent
to show that the microstates, compatible with the set of macroscopic
variables, suitable to describe the thermodynamic system at hand,
during an adiabatic evolution occupy phase-space regions of increasing
volumes \cite{lego}.  This fact guarantees the experimental
reproducibility of the process. On the other hand, as Jaynes writes
\cite{jaynes1}, ``Any really satisfactory demonstration of the second
law must therefore be based on a different approach than
coarse-graining [of $\rho$]'', since ``the decrease of $\overline{H}$
[a coarse-grained version of $H_G(\{\rho\})$] is due only to the
artificial coarse-graining operation and it cannot therefore have any
physical significance...''.

In the case of a dilute gas, a useful macroscopic variable to define
the thermodynamic state of the system, is the single particle
distribution function $f(\mathbf{q},\mathbf{p},t)$. In this case, and
only in this case, the logarithm of the volume occupied by the
compatible microstates is given by minus the Boltzmann $H$-function: $
H_B = \int f(\mathbf{q},\mathbf{p},t)\ln f(\mathbf{q},\mathbf{p},t)
\mathrm{d}\mathbf{q}\mathrm{d}\mathbf{p}$. So, for a dilute gas, while
the Gibbs entropy, as always, is expected to be constant, the
Boltzmann entropy takes the form: $ S_B = - k_{_B} \int
f(\mathbf{q},\mathbf{p},t)\ln f(\mathbf{q},\mathbf{p},t)
\mathrm{d}\mathbf{q}\mathrm{d}\mathbf{p}$ and is expected to increase,
as Lanford \cite{OL75} demonstrated to be true, in a suitable limit.
 
Most of the ideas discussed in the present paper are known. See, for
instance, \cite{lego} for the graining in phase space induced by a
graining in $\mu$-space and the choice of macro-variables, and for the
relation between the one-particle distribution function of interacting
and of noninteracting particle systems. Here, we express in a
quantitative form, for a specific model, the qualitative statements
made elsewhere, and we observe the existence of an intrinsic scale for
the graining in the $\mu$-space of interacting particle systems.

This paper is organized as follows. Section II is devoted to a brief
summary of some basic facts about Gibbs and Boltzmann entropy and the
link between deterministic chaos and statistical mechanics. In section
III we introduce a symplectic map simulating the time evolution of a
large system, consisting of weakly interacting chaotic subsystems, and
we show the behavior of the Boltzmann entropy $S_B$ vs time. The main
result is that, for interacting systems, $S_B$ is independent of the
details of the coarse-graining procedure in $\mu$-space, if the
graining is sufficiently fine, i.e.\ the existence of an intrinsic
graining scale in $\mu$-space.  The most remarkable fact is that some
numerical aspects (e.g. the slope of $S_B$ for large enough $t$)
basically depend only on the chaotic properties of the single
subsystem, but the presence of the coupling is absolutely necessary.
In section IV we show how one can interpret the results with a
mechanism similar to that used for the decoherence in the
semiclassical limit of quantum mechanics. Conclusions and perspectives
are in section V.

\section{On the Gibbs and Boltzmann entropy}

Let us consider a Hamiltonian system of $N$ particles, and the vector
$\mathbf{X}(t)$ which defines the microscopic state of the
system. Denoting by $\rho(\mathbf{X})d\mathbf{X}$ the probability for
the microscopic state to be found in the phase space volume
$d\mathbf{X}$, one defines the Gibbs entropy:
\begin{equation}\label{gibbs}
S_G(\{\rho\}) = - k_{_B} \int \rho(\mathbf{X})\ln
\rho(\mathbf{X})\mathrm{d} \mathbf{X}
\end{equation}
where $k_{_B}$ is the Boltzmann's constant.  Since the time evolution
$\mathbf{X}(0)\rightarrow\mathbf{X}(t)$ is ruled by the Hamilton
equations, the Liouville theorem implies that $S_G$ does not change in
time. However, a coarse graining of $\rho$ by cells of size $\Delta$ in
the $\Gamma$ space (the microscopic phase space) leads to an increase
of the coarse-grained Gibbs entropy\footnote{
It is worth recalling that $S_G(t,\Delta)$ is the
discretization of $S_G$, by cells of size $\Delta$, except for an additive 
term $ k_{_B} \ln \Delta$. At fixed $\Delta$, the term is constant 
and not relevant if one considers the entropy differences with 
respect to the initial values.}
\begin{equation}
S_G(t,\Delta) =- k_{_B} \sum_i p_\Delta(i,t)\ln p_\Delta(i,t) ~,
\end{equation}
where the coarse-grained probability $p_\Delta(i,t)$ is given by:
\begin{equation}
p_\Delta(i,t)=\int_{\Lambda_{i,\Delta}} \rho(\mathbf{X},t)
\mathrm{d} \mathbf{X} ~,
\end{equation}
and $\Lambda_{i,\Delta}$ is the cell in the space $\Gamma$ of linear
size $\Delta$ centered in the point $\mathbf{X}^{(i)}$. When the
system is chaotic, and the initial probability distribution is
supported over a small region of linear size $\sigma$, simple
arguments suggest that $S_G(t,\Delta)$ increases linearly in time,
after a short transient of length $t_\lambda$
\begin{equation}\label{sinaigrowth}
S_G(t,\Delta)-S_G(0,\Delta) \simeq \left \{ 
\begin{array}{cc}
0 & \quad t<t_\lambda\\
h_{KS}(t-t_\lambda) & \quad t\ge t_\lambda
\end{array}
\right.
\end{equation}
where $h_{KS}$ is the Kolmogorov-Sinai entropy of the system,
\begin{equation}
t_\lambda\sim\frac{1}{\lambda_1}\ln\left(\frac{\sigma}{\Delta}\right)
\end{equation}
and $\lambda_1$ is the first Lyapunov exponent of the Hamilton
equations. The prediction of Eq (\ref{sinaigrowth}) is not always
completely correct; actually, it is valid only when intermittency
effects are negligible \cite{palatella}. To make things simpler we
will assume the validity of Eq.(\ref{sinaigrowth}), as commonly done
in the presence of chaotic dynamics. Many consider Eq.\
(\ref{sinaigrowth}) as a proof of the deep connection between chaos
and irreversibility. Nevertheless, the result of
Eq.(\ref{sinaigrowth}) does not have a genuine thermodynamic
character. This can be understood as follows:
\begin{itemize} 
\item[$a$)] the quantity $S_G(t,\Delta)$
describes properties of the $\Gamma$ space that can not be computed
from a single-system measurement; 
\item[$b$)] the time increase of
$S_G(t,\Delta)$ depends on the coarse-graining procedure and this
appears as a non-ontological result;
\item[$c$)] the result of Eq.(\ref{sinaigrowth}) is valid also
for a generic, low-dimensional system - notice that one obtains the
same result even in the case of chaotic non-interacting particles.
\end{itemize}

Consider now the Boltzmann point of view, for a system of $N$ weakly
interacting particles. The one-body probability distribution function
$f(\mathbf{q},\mathbf{p},t)$, i.e. the probability density of finding
a particle in a given volume of the $\mu$ space (the single-particle
space) can be introduced without any reference to an ensemble of
macroscopically identical systems, each represented by a point in
$\Gamma$ space. In fact, take a single system made of a large number
$N$ of identical particles, consider the discrete distribution
\begin{equation}\label{singleparticle}
f(\mathbf{q},\mathbf{p},t) = \frac{1}{N} \sum\limits_{i=0}^{N}
\delta \left ( \mathbf{q} - \mathbf{q}_i(t) \right ) \delta (\mathbf{p} -
\mathbf{p}_i(t))
\end{equation}
and let $N$ grow. As the number $N$ of the particles of this {\em
unique} system tends to infinity, a non-singular one-particle
probability distribution (in $\mu$ space) must be constructed\footnote{This 
is needed for the Boltzmann entropy to exist, like
a non-singular distribution in $\Gamma$ space is required for the
Gibbs entropy to exista} 
Using Eq.(\ref{singleparticle}), and assuming that the
physical space is $d$-dimensional, this can be done introducing a cell
size $\Delta$ in $\mu$ space, such that $N \gg \Delta^{-2d}$, i.e.\
that there is a statistically relevant number of particles in each
cell. Then, define the one-particle coarse grained distribution:
\begin{equation}\label{singlemesh}
f_{\Delta}(\mathbf{q}^{(j)},\mathbf{p}^{(k)},t) = \frac{1}{N}
\sum\limits_{i=0}^{N} \Theta \left (1 - \frac{2| \mathbf{q}^{(j)} -
\mathbf{q}_i(t) |}{\Delta} \right ) \Theta \left (1 -
\frac{2|\mathbf{p}^{(k)} - \mathbf{p}_i(t)|}{\Delta} \right )
\end{equation}
where $\Theta(z)$ is the Heaviside step function and
$\mathbf{q}^{(j)}$, $\mathbf{p}^{(k)}$ are the coordinates of the
center of each cell $C_{jk}$ having linear size $\Delta$, and volume
$\Delta^{2d}$, in the appropriate units. The Boltzmann entropy is then
defined by
\begin{equation}\label{bolentr}
S_B(t)=- k_{_B} \int f(\mathbf{q},\mathbf{p},t)\ln f(\mathbf{q},\mathbf{p},t)
\mathrm{d}\mathbf{q}\mathrm{d}\mathbf{p}
\end{equation}
where $f$ is the regular $\mu$-space probability distribution,
obtained in the $N \to \infty$, $\Delta \to 0$ limit of
Eq.(\ref{singlemesh}). The Boltzmann entropy, as defined in
Eq.(\ref{bolentr}), is a natural candidate for the description of
dilute systems, for which it represents the logarithm of the volume
occupied by the macrostate in the $\Gamma$ space:
\begin{equation}
S_B= k_{_B} \log \Delta\Gamma.
\end{equation}
Moreover, for dilute systems under the hypothesis of molecular chaos
the celebrated Boltzmann's H-theorem holds:
\begin{equation}
\label{growth}
\frac{d S_B}{dt}\ge 0.
\end{equation}
The validity of the molecular chaos hypothesis has been demonstrated
for the class of dilute systems in the Grad limit, where
$N\rightarrow\infty$ and the interaction range goes to zero in order
to keep the total cross section constant \cite{OL75,IP-PI}.

Some textbooks try to connect the two main approaches noticing that in
dilute systems:
\begin{equation}
\label{dens-prod}
\rho(\mathbf{X},t)\simeq \prod\limits_{j=1}^N f(\mathbf{q}_j,\mathbf{p}_j,t)
\end{equation}
which implies that $S_G\simeq N S_B$.
This attempt, however, is only partially justified, as there are at least
two important conceptual differences between these approaches:

\begin{itemize}
\item The Gibbs point of view is based on the ensemble, i.e.\ on an
abstract collection of macroscopically identical systems, and does not
depend on the number of particles of which each system is
made. Differently, Boltzmann's approach does not require an ensemble
of copies of the same system, but needs $N\gg1$, in order to compute
$f(\mathbf{q},\mathbf{p},t)$ for the single system.

\item 
The Gibbs entropy deals with the $\Gamma$ space, and necessitates a
coarse graining procedure in order to avoid the consequences of the
Liouville theorem, and to grow during an irreversible evolution. 
In the Boltzmann approach, the entropy can grow despite the Liouville 
theorem, and the graining of the $\mu$-space is only introduced 
to deal with a smooth distribution. 

\end{itemize}
In addition, the validity of (\ref{dens-prod}) must be interpreted
\textit{cum grano salis}, otherwise, because of (\ref{growth}), one would
erroneously infer that the Gibbs entropy (\ref{gibbs}) grows.

In the following, we will show that for long enough times, and
non-vanishing interaction, the growth of $S_B$ does not depend on the
cell size, which is not the case for non-interacting systems.

\section{The Boltzmann entropy of a chaotic system}
\subsection{The discrete time model}

In order to discuss the points mentioned above in a concrete fashion,
let us start with a system consisting of $N\gg1$ non-interacting
particles moving in a periodic array of fixed convex scatterers, with
which they collide elastically.

The position of the scatterers should avoid the presence of
collisionless trajectories, i.e.\ the horizon should be finite. It is
well established that such a system, commonly known as the Lorentz 
gas\footnote{In Lorentz's original model, the moving particles were 
considered in thermal
equilibrium with the scatterers, which is impossible to achieve
without energy exchanges between scatterers and particles, as in the
present model.}
is chaotic and displays asymptotic diffusion. The Gibbs
entropy of such a system obeys $S_G(t)=S_G(0)$, while the
coarse-grained entropy increases linearly with $t$ for $t>t_\lambda$
(before saturation).  It is easy to see that $h_{KS}=N h_1$, where
$h_1$ is the Kolmogorov-Sinai entropy of a single particle, and $S_B =
(S_G)/N = $ constant, because the particles are independent, and the
probability distribution in phase space factorizes in $N$ identical
terms.

The situation is quite different in the case of interacting particles,
such as the case of the generalization of the Lorentz gas given in
Ref.\cite{BGG}.  There, the particles are not point-like, but have a
finite size, and therefore they collide not only with the scatterers,
but also among themselves.

Unfortunately, the study of such a system, for a sufficiently large
number of particles, is very expensive from a computational point of
view. However, a reasonable substitute for such a system, which shares
its main features, can be given in terms of symplectic maps. For
instance, one can consider a two-dimensional map, with one
``coordinate'' and one ``momentum'', in place of each particle of the
generalized Lorentz gas, and one can introduce a form of interaction
among these ``particles''. Thus we require that:
\begin{itemize}

\item in the absence of interactions among the ``particles'', the 
single-particle dynamics in the corresponding $\mu$ space be chaotic 
and volume preserving;

\item in the presence of interactions among particles,
the dynamics of the whole system, described by the vector
$\mathbf{X}=(\mathbf{Q},\mathbf{P})$, $\mathbf{Q}=(q_1\dots
q_n)$, $\mathbf{P}=(p_1\dots p_n)$, be symplectic and volume 
preserving in the $\Gamma$ space.

\end{itemize}
The resulting model will be numerically easier to handle, still having
some important properties of the particle system. In particular, the
dynamics of the interacting case will not be volume preserving in the
$\mu$ space. To this aim, we introduce the symplectic map:
\begin{equation}\label{simpldyn}
\left \{
\begin{array}{ccc}
q_i & = & \partial G(\mathbf{Q'},\mathbf{P})/\partial p_i \quad \mod \
1 \\ p'_i & = & \partial G(\mathbf{Q'},\mathbf{P})/\partial
q'_i \quad \mod \ 1
\end{array}\right . 
\end{equation}
whose generating function $G(\mathbf{Q'},\mathbf{P})$ is defined by:
\begin{eqnarray}\label{generfun}
G(\mathbf{Q'},\mathbf{P}) = \mathbf{Q'}\mathbf{P}-
\frac{|\mathbf{P}|^2}{2}- \frac{k}{2\pi} \sum\limits_{i=0}^N
\sum\limits_{j=0}^{N_{S}} \cos \left [2\pi( q'_i - Y_{j}) \right] -\nonumber\\
-\frac{\epsilon}{4\pi} \sum\limits_{i=0}^N \sum\limits_{n=-M/2}^{M/2} \cos
\left[2\pi ( q'_i - q'_{i+n} )\right ]
\end{eqnarray}
with $q_i, p_i\in[0,1]$. $N_S$ is the number of fixed ``obstacles''
having positions $Y_j$, which play the role of the convex scatterers
in the Lorentz gas, and $N$ is the number of ``particles''. The
parameters $k$ and $\epsilon$ represent the interaction strength
between particles and obstacles and among particles respectively.  If
$k=\epsilon=0$ one has free particles.  The functional form of the
generating function is reminiscent of the standard map, which is a
paragon of symplectic dynamics. The boundary conditions on the
variables are periodic, and the form of the interactions does not
present discontinuities at the boundaries.  In order to make the
numerical simulations faster, we assume that each particle interacts
only with a limited number $M$ of other particles.

Substituting Eq.(\ref{generfun}) into (\ref{simpldyn}), one finds:
\begin{equation}\label{sistema}
\left \{
\begin{array}{lll}
q'_i = & q_i + p_i \mod \: 1 &\\ p'_i = & p_i + k
\sum\limits_{j=0}^{N_{S}} \sin[ 2\pi \left ( q'_i - Y_{j} \right )] +
\epsilon \sum\limits_{n=-\frac{M}{2}}^{\frac{M}{2}} \sin [2 \pi \left ( q'_i -
q'_{i+n} \right ) ]\  \mod \: 1&
\end{array}
\right.
\end{equation}
Since the system is symplectic, the dynamics described by the points
$(\mathbf{Q},\mathbf{P})$ will preserve volumes in phase space. 

\subsection{Numerical results}

In the following, we calculate the Boltzmann single-particle
distribution for a given cell size (cf.\ Eq.(\ref{singlemesh})), as a
function of time. Then, we study the growth of the corresponding
Boltzmann entropy with time, defined by
\begin{equation}\label{boltzmanngrowth}
S_B(t,\Delta)=- k_{_B} \sum_{j,k} f_{\Delta}(q^{(j)},p^{(k)},t) \log
f_{\Delta}(q^{(j)},p^{(k)},t)
\end{equation}
by varying the interaction strength $\epsilon$ and the cell size
$\Delta$.  Note that, in our numerical computations, where we set
$k_{_B} =1$ for convenience, there are no {\em a priori} assumptions
such as, for example, the hypothesis of molecular chaos. In other
words, the quantity $f_\Delta(q, p,t)$ evolves according to the exact
dynamics.

Let us comment on the use of the entropy defined in
Eq.(\ref{boltzmanngrowth}) for our dynamical system. As discussed in
the Introduction, this expression is correct (in the sense that it
measures the interesting volume of phase space) only in the case of
dilute systems. For systems, where the potential energy is not a tiny
fraction of the total, it has been proposed \cite{lego,lebPRL} that a
similar recipe may still be used, with the prescription to count only
microstates corresponding to a fixed total energy $E$. Since we are
considering the case of weakly interacting subsystems, the function
$f(q, p,t)$ is able to properly describe the macrostates of the system

\begin{figure}[!ht]
\begin{center}
\includegraphics[width=9 cm]{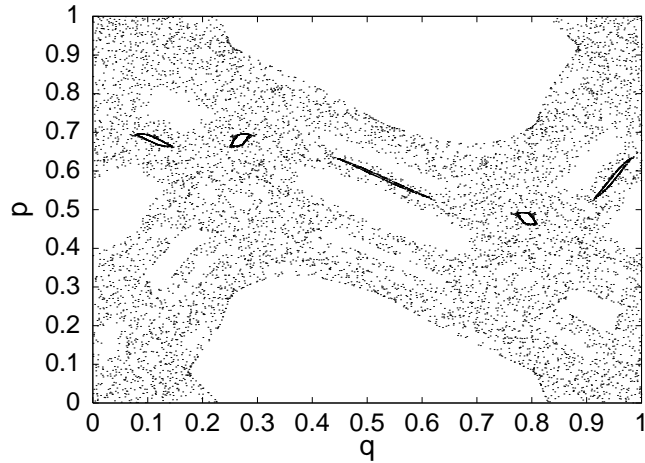}
\end{center}
\caption{\label{fig1} A trajectory generated by $10^4$ map
iterations in the phase space of a single particle, with $N_S=10^3$,
$k=0.017$,$\epsilon=0$.}
\end{figure}

First of all, we choose the number of obstacles $N_S$ and the
parameter $k$ (both related to the single-particle chaotic behavior)
in such a way that a) the Lyapunov exponent of the single particle
dynamics is not too large and b) there are no KAM tori, of the kind
which constitute barriers for the transport. One possible choice,
which realizes this requirement is $N_S=10^3$ and $k=0.017$.  The
obstacles positions are selected at random, with a uniform p.d.f. The
result is that the Lyapunov exponent is $\lambda\approx 0.162$. An
example of a trajectory in the 1-particle $(q,p)$ space is shown in
Fig.\ref{fig1}.

For the number of particles, we chose $N=10^7$, after checking that
finite size effects are negligible for this $N$, and we let the
simulations run for different values of $\epsilon$. We checked that
the results basically do not change when the number of particles is
reduced by a factor $10$.  As initial non-equilibrium condition, we
take a cloud of points distributed according to a Gaussian of r.m.s.d
$\sigma=0.01$, having the point $(q,p)=(1/4,1/2)$ as a center (we
checked that this point is far enough from the surviving regular
islands, see Fig. \ref{fig1}). At each time, we compute the
differences between the entropy and its initial value at several
resolutions $\Delta$ :
\begin{equation}
\delta S(t,\Delta)=S(t,\Delta)-S(0,\Delta).
\end{equation}

\begin{figure}[!ht]
\begin{center}
\includegraphics[width=9 cm]{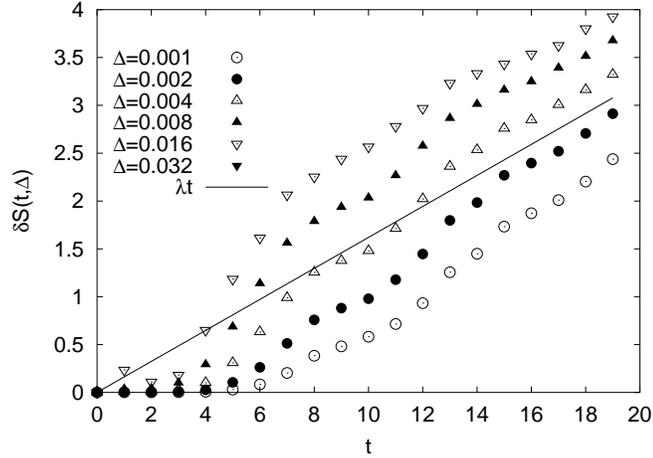}
\end{center}
\caption{\label{fig2a} $\delta S(t,\Delta)$ with $\epsilon=0$
(non-interacting particles) as a function of $t$ for different values
of $\Delta$. The slope of the straight line equals Lyapunov exponent.}
\end{figure}

\begin{figure}[!ht]
\includegraphics[width=5.8 cm]{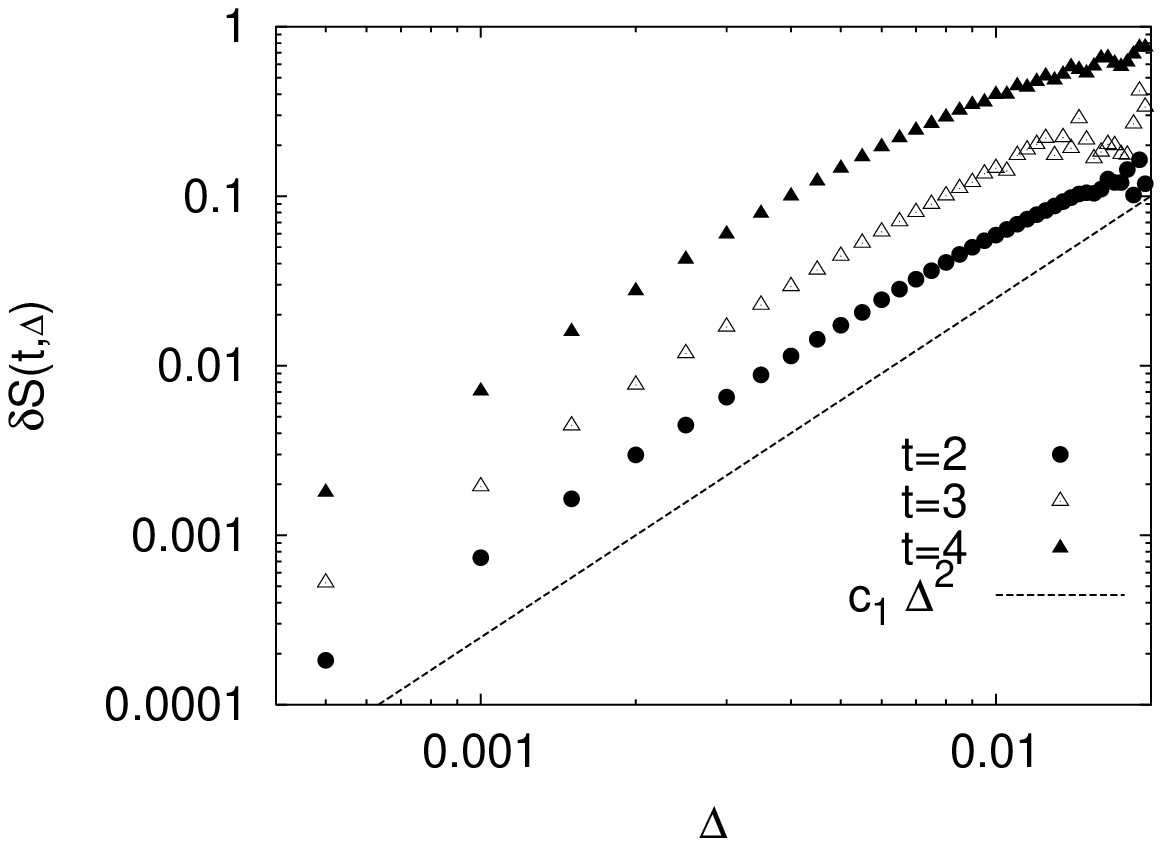}
\includegraphics[width=5.8 cm]{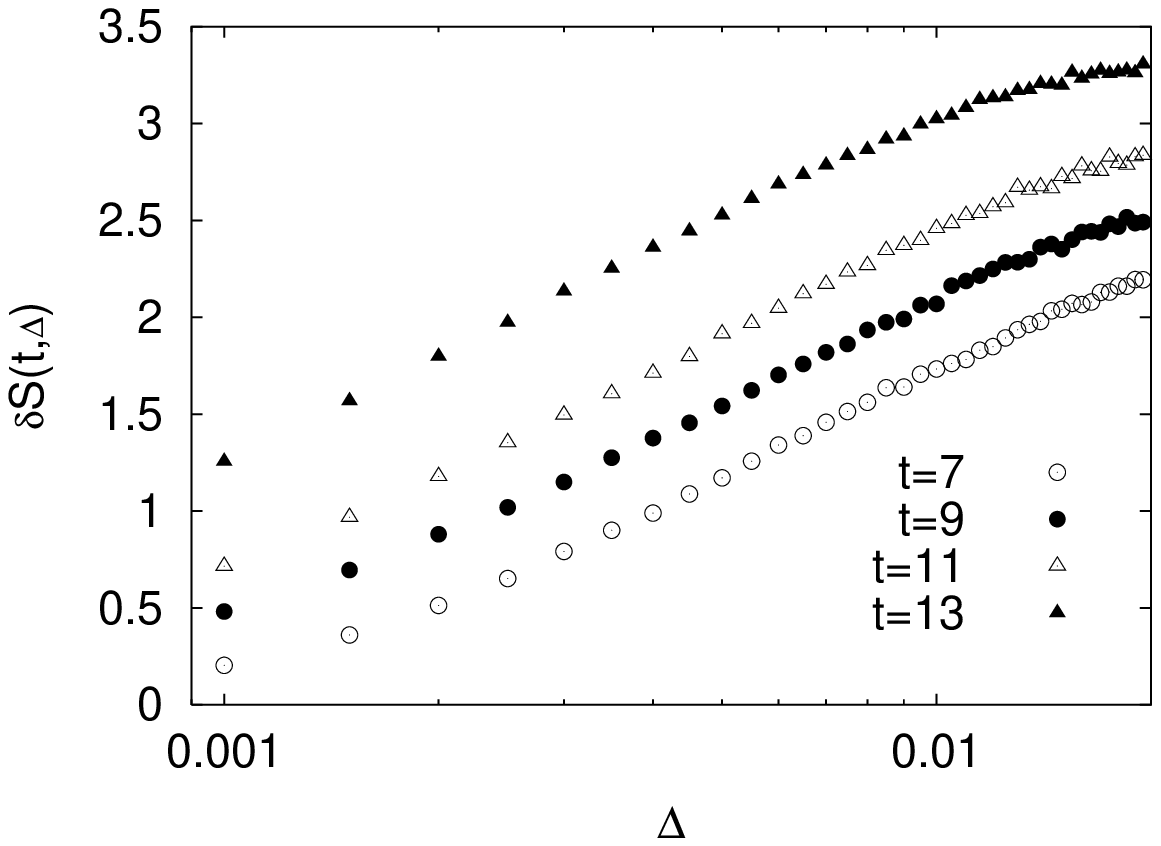}
\caption{\label{fig2b} $\delta S_B(t,\Delta)$ with $\epsilon=0$
(non-interacting particles) as a function of $\Delta$, for small
values of $t$ (left) and large values of $t$ (right). In the left
panel, the dotted line shows the expected behavior $\delta
S_B(t,\Delta) \propto \Delta^2$, while the logarithmic behavior is
clearly visible in the right panel (notice the log-linear scale).}
\end{figure}

We begin with the study of the $\epsilon=0$ case. The entropy growth shown
in Fig.\ref{fig2a} is only due to the discretization procedure, since
the equation ruling the evolution of $f(q,p,t)$ obeys the Liouville
theorem. This means that the ``true'' Boltzmann entropy for
$\Delta\rightarrow 0$ is constant in time.  As shown in
Fig.\ref{fig2a}, the curves of the entropy differences as functions of
time stay constant up to a time $t_\lambda$ depending on $\Delta$.
After this transient, the slope of $\delta S(t,\Delta)$ is practically
the same for all the curves and is approximately given by $h_{KS}$
(see Eq.\ref{sinaigrowth}). Looking at the curves of the entropy
differences as a function of $\Delta$ (see Fig.\ref{fig2b}), it is
possible to extrapolate the behavior for $\Delta\rightarrow 0$: far
from the saturation (i.e. for small times) and for $\Delta$ not too
large, these curves are well fitted by a power law:
\begin{equation}\label{quadrato}
\delta S_B(t, \Delta)\propto \Delta^2.
\end{equation}
This result suggests that the relevant parameter for understanding the
finite resolution behavior of the entropy differences is the cell area
$\Delta^2$; moreover, these differences go correctly to zero when
$\Delta\rightarrow 0$.  For $t$ larger than $t_\lambda$, one observes
\begin{equation}\label{logaritmo}
\delta S_B(t, \Delta) = a \log(\Delta) + b.
\end{equation}
Of course, the behavior of Eq. (\ref{quadrato}) and (\ref{logaritmo})
is consistent with Eq.(\ref{sinaigrowth}), taking into account that
the rate of entropy growth after $t_\lambda$ is generally
different from the Lyapunov exponent, leading to $a\neq 1$ \cite{palatella}.

\begin{figure}[!ht]
\begin{center}
\includegraphics[width=9 cm]{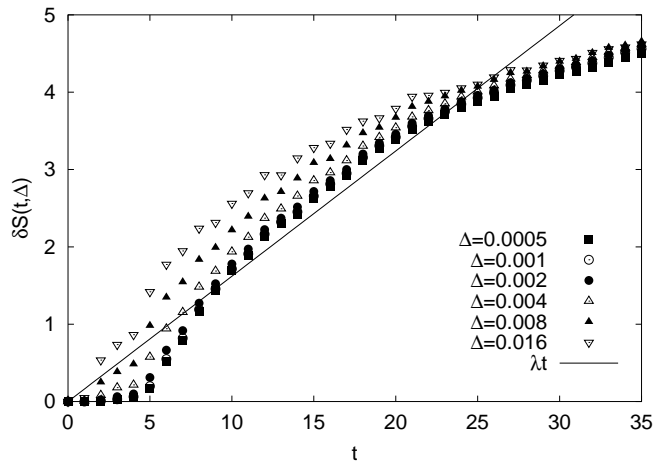}
\end{center}
\caption{\label{fig3a}  $\delta S_B(t,\Delta)$ with $\epsilon=10^{-4}$
as a function of $t$ for different values of $\Delta$. The straight
line slope equals the Lyapunov exponent.}
\end{figure}

\begin{figure}
\includegraphics[width=5.8 cm]{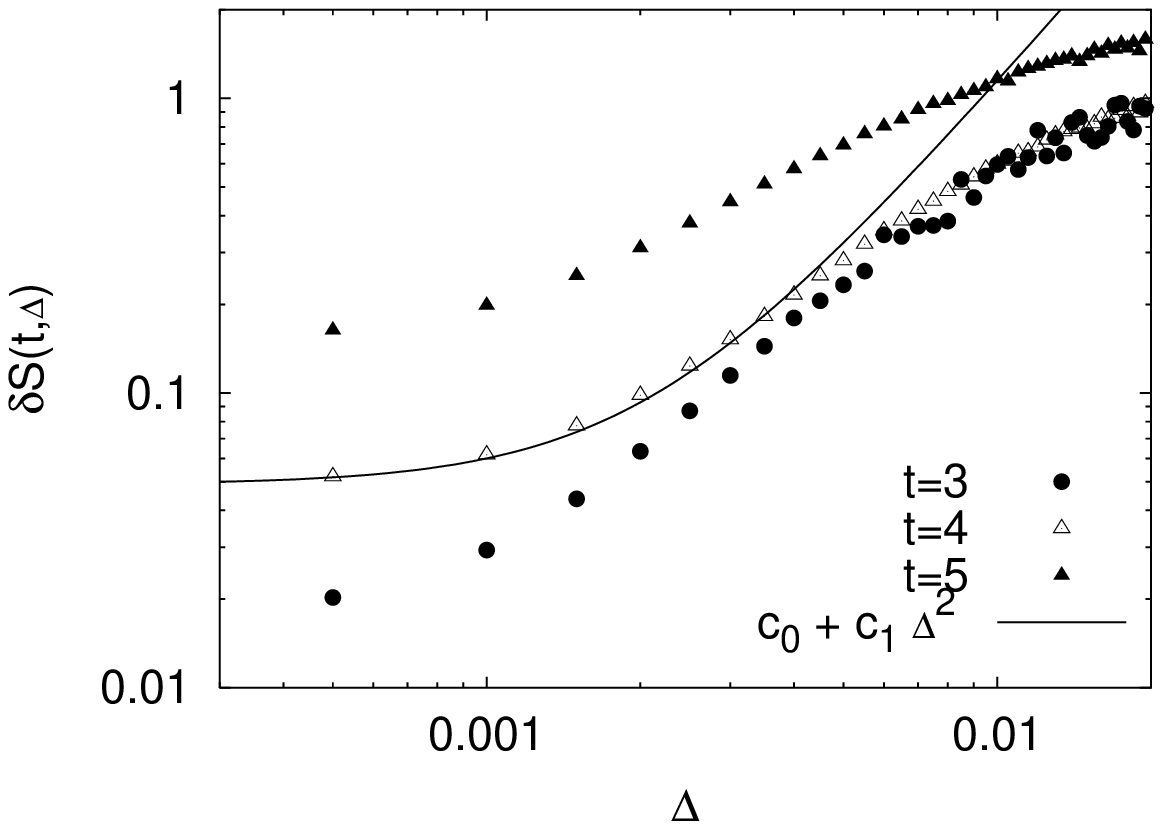}
\includegraphics[width=5.8 cm]{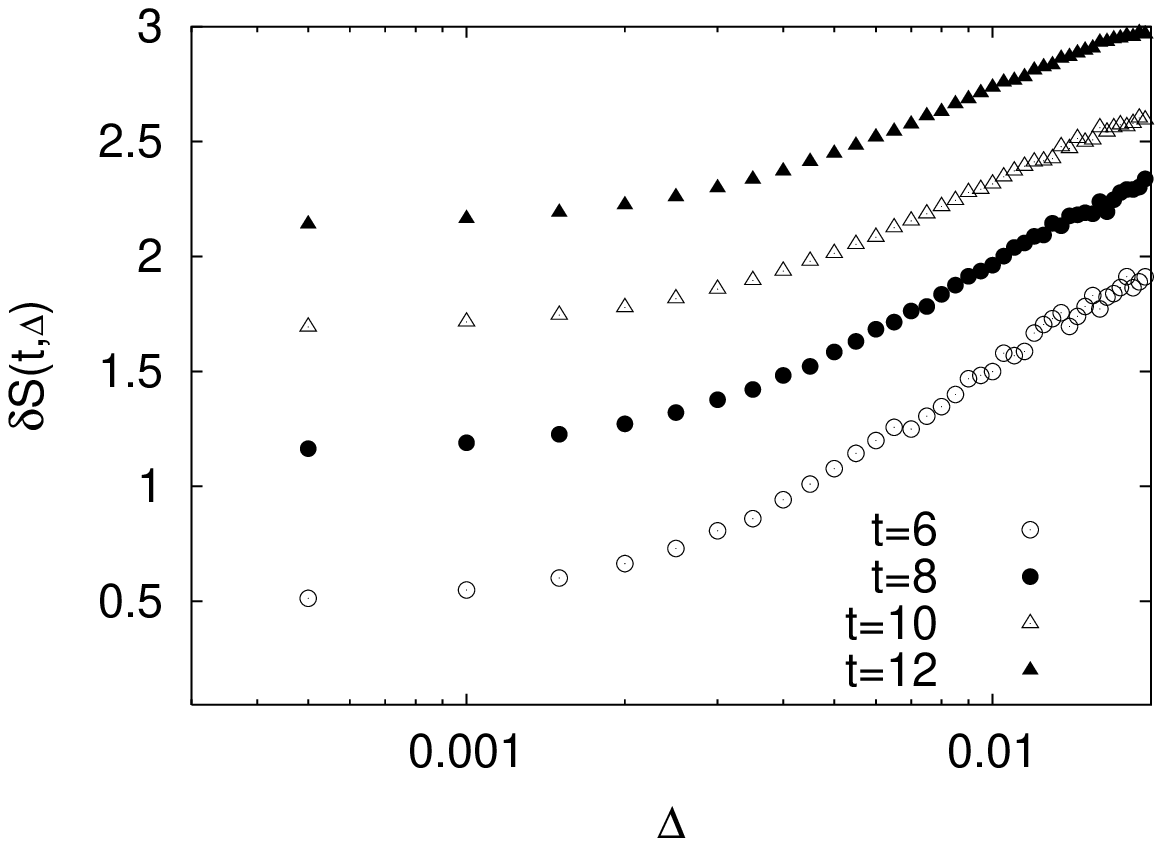}
\caption{\label{fig3b} $\delta S_B(t,\Delta)$ with $\epsilon=10^{-4}$
as a function of $\Delta$ for small values of $t$ (left panel) and
large values of $t$ (right panel). In the left panel the line shows
the expected behavior $c_0 + c_1 \Delta^2$ while in the right panel
$\delta S_B(t,\Delta)$ shows a weak dependence on $\Delta$ for $\Delta
\rightarrow 0$.}
\end{figure}

We consider now the ``interacting'' case, i.e. $\epsilon>0$. Figure
\ref{fig3a} and \ref{fig3b} show the curves of $\delta S_B(t,\Delta)$
as a function of $t$ and $\Delta$.  In this case, the entropy curves
as a function of $\Delta$ do not extrapolate anymore to zero (see
Fig.\ \ref{fig3b}); these curves, for small (fixed) times, are well
fitted by a polynomial like:
\begin{equation}
\label{deltaS}
\delta S_B(t,\Delta)\approx c_0 +c_1 \Delta^2.
\end{equation}

After a characteristic time depending on $\epsilon$,
$t_*(\epsilon,\Delta)$, the entropy shows just a weak (logarithmic)
dependence on $\Delta$ and correctly extrapolates to a finite value
when $\Delta\rightarrow 0$.

\begin{figure}[!ht]
\begin{center}
\includegraphics[width=9 cm]{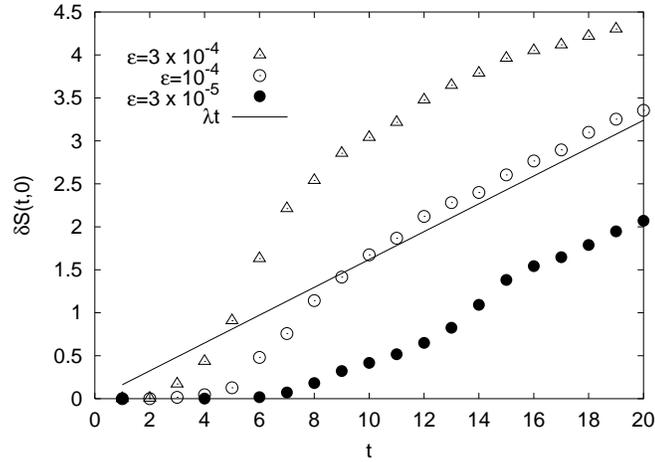}
\end{center}
\caption{\label{fig4} Extrapolation for $\Delta\rightarrow 0$ of the
curves $\delta S_B(t,\Delta)$ as a function of $t$ for various values of
$\epsilon$.}
\end{figure}

\begin{figure}[!ht]
\begin{center}
\includegraphics[width=9 cm]{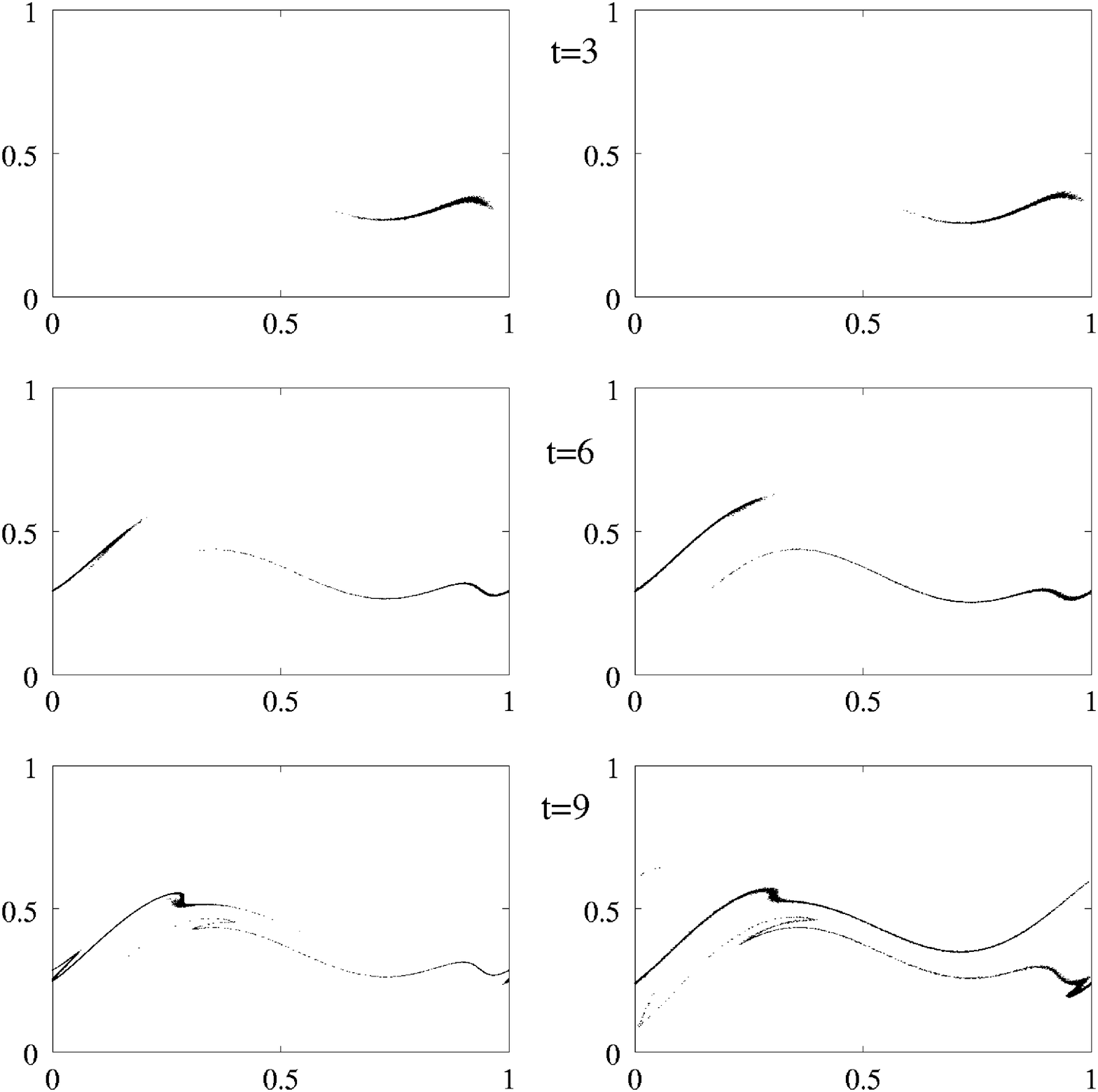}
\end{center}
\caption{\label{fig5}Snapshots of the evolution of the system in the
single-particle projection of the phase space in the non-interacting
case (left), and in the interacting case with $\epsilon=10^{-4}$
(right) and $M=100$, at increasing times from top to bottom, as indicated
in the figure.}
\end{figure}

Let us now summarize and comment the previous results:
\begin{itemize}

\item[a)] for non-interacting systems ($\epsilon = 0$), the growth of
$\delta S_B(t,\Delta)$ reflects the properties of the observation
tools, i.e. $\delta S_B\simeq0$ for $t\lesssim t_\lambda(\Delta)$ and
$\delta S_B \simeq \lambda (t-t_\lambda(\Delta))$ for $t\gtrsim
t_\lambda(\Delta)$, has a kind of ``subjective'' character.  Since
$t_\lambda$ increases as $\Delta$ decreases, not only does the value
of the entropy depend on the coarse graining, but the entropy growth
depends (for ``small'' $t$) on the resolution scale, as noticed in
\cite{RC-CR}.

\item[b)] For weakly interacting systems, there is an effective cell
size $\Delta_*(\epsilon,\lambda)$, such that if
$\Delta<\Delta_*(\epsilon,\lambda)$ the value of $\delta
S_B(t,\Delta)$ does not depend on $\Delta$. Here, the entropy growth
is an objective property, meaning that the limit for
$\Delta\rightarrow0$ of $\delta S_B(t,\Delta)$ exists, is finite,
hence is an intrinsic property of the system (cf.\ Fig.\ref{fig3b}).

\item[c)] The role of chaos in the limit of vanishing coupling is
relevant, i.e. the slope of $\delta S_B(t,\Delta)$, for $t$ large
enough, is given by the Lyapunov exponent, but the existence of an
effective cell size $\Delta_*(\epsilon,\lambda)$ and the corresponding
$t_*(\epsilon, \lambda)$ depends on the coupling strength $\epsilon$,
and on $\lambda$.

\item[d)] In the above procedure, i.e. in the evaluation of
Eqs. (\ref{singlemesh}) and (\ref{boltzmanngrowth}), there are no
assumptions like the hypothesis of molecular chaos or of system's
dilution. From a mathematical point of view, we can define
$S_B(t,\Delta)$ in Eq.(\ref{boltzmanngrowth}) in full generality.
However, we consider only the weakly interacting limit of small
$\epsilon$ for the physical reason that only in such a case does
$f(q,p,t)$ afford an appropriate thermodynamic meaning.

\item[e)] For small values of $\epsilon$, the time evolution of
$f(q,p,t)$ is different from the case $\epsilon=0$ only on very small
graining scales; in other words, the coupling is necessary for the
``genuine'' growth of the entropy, but it does not have any dramatic
effect on $f(q,p,t)$ at scales $\Delta\lesssim\Delta_*$. Indeed, as
shown in Fig.\ref{fig5}, the noninteracting and the weakly interacting
cases do not appear to be so different, in terms of the
single-particle phase space distribution.

\end{itemize}

\section{Interpretation of the results}

The results of the previous section suggest the following
interpretation: since the number of particles is large, one can expect
that the effect of the interaction on each particle may be reasonably
described by some kind of thermal bath.  The single particle dynamics
can then be mimicked by chaotic dynamics (corresponding to the
symplectic map of Eq.(\ref{sistema}) with $\epsilon=0$) coupled to a
noise term whose strength is $O(\epsilon)$:
\begin{equation}\label{noisy}
\left \{
\begin{array}{lll}
q_i(t+1) &= q_i(t) + p_i(t) & \quad \mod \: 1\\ p_i(t+1) &= p_i(t) + k
\sum\limits_j \sin \left[2 \pi (q_i(t+1)-Y_j) \right] + \sqrt{2D} 
\eta_i(t) & \quad
\mod \: 1
\end{array}
\right .
\end{equation}
where $\eta_i(t)$'s are i.i.d.\ Gaussian variables with zero mean and
unitary variance, i.e.
\begin{equation}
\langle \eta_i(t) \rangle = 0, \quad \langle \eta_i(t) \eta_j(t')
\rangle = \delta_{t,t'} \delta_{i,j}.
\end{equation}
With this approximation, one basically assumes that $f(q,p,t)$ evolves
according to a discrete time Fokker-Planck equation.  Supposing that
each particle gives an uncorrelated contribution to the noise term,
one can roughly estimate the diffusion coefficient $D$ as
$M\epsilon^2/4$.  This heuristic estimate is well supported by
numerical simulations of (\ref{sistema}): the quantity $\delta
S_B(t,\Delta)$ practically does not change at varying $M$ and
$\epsilon$, keeping $M \epsilon^2$ constant.

In this framework, one can introduce a characteristic time $t_{c}$,
defined as the time in which the scale of the noise induced diffusion
reaches the smallest scale originated by the deterministic chaotic
dynamics \cite{pattanayak}.  This definition of $t_c$ would correspond
to $t_*(\epsilon,\lambda)$ introduced above. Consequently, noting that
the typical lengths due to noise and to chaos behave as
$\sqrt{M\epsilon^2t/2}$ and $\sigma\exp(-\lambda t)$, respectively,
the time $t_{c}$ may be estimated as the solution of the following
transcendent equation:
\begin{equation}\label{zurpaz}
\epsilon \sqrt{M t_{c}/2} = \sigma \exp(-\lambda t_{c})
\end{equation}
which holds on the spatial scales already
reached by the diffusion process:

\begin{equation}
\epsilon \sqrt{M t/2} > \Delta  ~,
\end{equation}
beyond which the value of the entropy still depends on the size of
$\Delta$ (i.e.\ the curves $S_B(\Delta)$ display the behavior
$S_B(t,\Delta)\sim \Delta^2$).  For example, in the case with
$\epsilon=10^{-4}$, $\sigma=0.01$, $M=100$ and $\lambda=0.162$, one
obtains $t_{c} \simeq 9$, and, in agreement with our interpretation,
for $t>9$, all the curves in Fig.\ref{fig4} present the same slope
compatible with the Lyapunov exponent $\lambda$.

\begin{figure}[!ht]
\begin{center}
\includegraphics[width=9 cm]{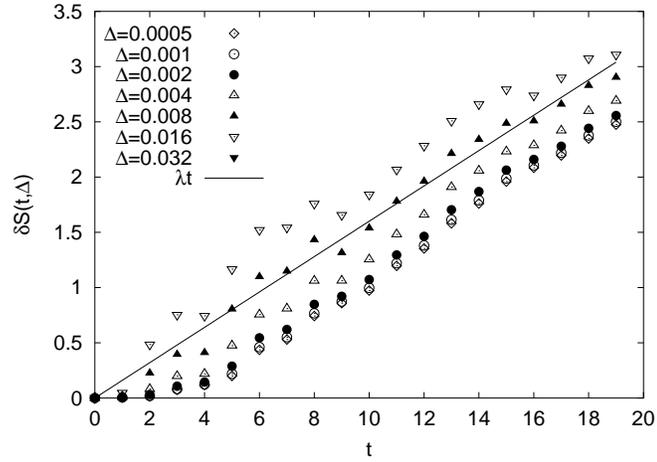}
\end{center}
\caption{\label{fig6} Numerical simulation of $N=10^7$ independent
particles evolving according to (\ref{noisy}) with $\epsilon_{eq} =
10^{-4}$.  Notice the similar qualitative behavior observed in
Fig.(\ref{fig3a}).}
\end{figure}

\begin{figure}[!ht]
\begin{center}
\includegraphics[width=6 cm,angle=-90]{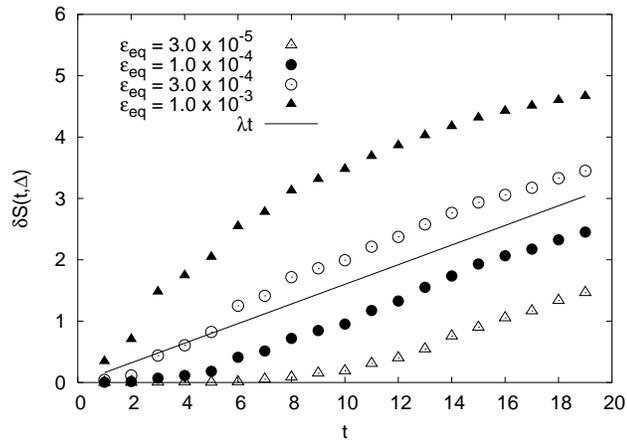}
\end{center}
\caption{\label{fig7} Extrapolation for $\Delta\rightarrow 0$ of the
curves $\delta S_B(t,\Delta)$ as a function of $t$ for various values of
$\epsilon_{eq}$. Notice that the extrapolated curves are qualitatively
similar to the interacting case (cf.\ Fig.\ \ref{fig4}).}
\end{figure}

As a numerical check of the consistency of this approach, we studied
system (\ref{noisy}). The results with a given $D$ should be compared
with those of the deterministic system of Eq.(\ref{sistema}) with
\begin{equation}
\epsilon=\epsilon_{eq} \equiv 2 \sqrt{\frac{D}{M}}.
\end{equation}
The results, shown in Fig.\ref{fig6} and \ref{fig7}, are qualitatively
similar to the deterministic interacting case (\ref{sistema}),
confirming the validity of our approach.

A similar reasoning leads to the decoherence mechanism proposed by
Zurek and Paz \cite{zurek} for the semiclassical limit of quantum
mechanics. We note that rather subtle conceptual points are present in
the decoherence process for the semiclassical limit.  This is so
because two theories are involved (classical and quantum mechanics)
with very different ontological status (deterministic and
non-deterministic, respectively). In our problem one has just a
technical aspect: roughly speaking, one mimics the first equation of
the BBGKY hierarchy of a diluted system, consisting of weakly
interacting chaotic particles, with a suitable Fokker-Planck equation.

\section{Conclusions and perspectives}

We have studied a system made of weakly coupled chaotic subsystems,
which can be considered as a model of weakly interacting particles in
an environment with convex obstacles.  In the non interacting limit
the behavior of the Boltzmann entropy depends strongly on the
coarse-graining resolution (\textit{i.e.}  the cell size $\Delta$) and
therefore it cannot be considered an intrinsic property of the
system. By contrast, in the weakly interacting case the behavior of
the Boltzmann entropy becomes independent of the observation scale,
assuming an objective character: for $\Delta$ small enough,
\textit{i.e.} smaller than $\Delta _{*} (\epsilon, \lambda)$, one
observes a well defined shape of $\delta S_B(t)$ $vs$ $t$. A
remarkable fact is that $\delta S_B(t)$, for $t \gtrsim t_* (\epsilon,
\lambda)$, increases roughly linearly, with a slope given by the
Kolmogorov-Sinai entropy of the single (non interacting) chaotic
system.  Summarizing, the interaction is necessary in order to have an
effective cell size $\Delta_*$, thus changing the time behavior of the
Boltzmann entropy from ``subjective'' to ``objective'', while other
numerical aspects, like the slope of $\delta S_B(t)$for $t \gtrsim
t_*$, are basically determined by the degree of chaos in the single
subsystems.  In addition, the effect of the weak coupling among the
chaotic subsystems can be successfully modeled with a noisy term, and
this allows us to estimate the value of $t_* (\epsilon, \lambda)$.

Let us now comment on the relevance of these results for the case of
noninteracting particles, in a host environment made of very heavy
particles.  If the ``obstacles'' were not infinitely massive, they
would exchange energy with the independent particles, like the case of
photons in a black body cavity. The photons do not interact with each
other, but interact with the walls, reach a thermal equilibrium with
them, and acquire a temperature for themselves. A similar behavior has
been obtained in the Lorentz-like model of Ref.\cite{MMLL}. In our
framework, this would amount to set the interaction $\epsilon$ to
zero, and to switch on an interaction between particles and obstacles
which now are allowed to move. Denoting with $Y_J$ and $W_J$ the
coordinate and the momentum of the $J$-th obstacle, one can introduce
a symplectic dynamics, which generalizes (\ref{simpldyn}) introducing
a suitable interaction among the ``light'' particles and the heavy
obstacles. Of course now the $\Gamma$ space is given by
$(\mathbf{Q},\mathbf{Y};\mathbf{P},\mathbf{W})$.  Then, the ``light''
particles would be indirectly coupled with each other, as in
\cite{MMLL}: indeed, the heavy particles will play the role of
interaction carrier particles and essentially we would fall back in
the case considered in this paper. Clearly, an interaction carried by
heavy particles will be very weak and will involve very small scales,
which are difficult to observe in numerical simulations; nevertheless,
there are no reasons to expect any conceptual difference from the
scenario described in the present paper, which is consistent with the
results of Ref.\cite{MMLL}.

Of course, our results do not rule out noninteracting particle systems
from the class of physically relevant models. In fact, they can be
quite useful in describing highly rarefied gases, such as those of
high vacuum pumps, or in the upper layers of the atmosphere. Recently,
they have also proven to be useful in the description of transport in
microporous membranes \cite{JR05}, which are of a strong technological
interest.  Therefore, credit must be given to those who have analyzed,
and popularized these models, in the framework of nonequilibrium
statistical mechanics (cf.\ \cite{PG98,TVM01} and references therein).
Similarly, we do not exclude that some version of the coarse-grained
Gibbs entropies may serve as useful characterizations of the state of
certain systems. Our results indicate that such systems, and such
definitions of entropy, are not appropriate to understand, from a
theoretical point of view, what happens in thermodynamic systems where
the macroscopic evolutions toward equilibrium involve all the relevant
microscopic degrees of freedom.

\section*{Acknowledgements}
The authors are grateful to J.R.\ Dorfman, P.\ Gaspard, T.\ Gilbert,
O.G.\ Jepps, J.L.\ Lebowitz, T.\ T\'el and J.\ Vollmer for their
extremely useful comments on a preliminary version of this paper.

\end{document}